\documentclass{hep99}
\usepackage{epsfig}
\usepackage{rotating}

\newcommand{\lsim}{\mbox{$\stackrel{\scriptstyle <}
{\scriptstyle \sim}$}}
\newcommand{\gsim}{\mbox{$\stackrel{\scriptstyle >}
{\scriptstyle \sim}$}}
\newcommand{\Msol}{\mbox{M$_{\odot}$}}
\begin{document}

\title{Astroparticle Physics}
\author{Norbert Magnussen}

\address{ Fachbereich Physik, Universit\"at Wuppertal, D-42097 Wuppertal, Germany\\[3pt]
E-mail: {\tt magnus@wpos7.physik.uni-wuppertal.de}}

\abstract{This article\dag\ reviews some recent developments in 
Astroparticle Physics. Due to the extension of the field only part
of the results and developments can be covered. The status of the search
for Dark Matter, some recent results on Cosmic Rays and Gamma Ray Astronomy
and the status of Neutrino Astronomy are presented. %\\[6pt]
}

\maketitle
\fntext{\dag}{Invited review at EPS-HEP '99, July 1999, Tampere, Finland.}

\section{Introduction}

Astroparticle Physics is rapidly growing into a diversified field of
observational and phenomenological physics, addressing questions
ranging from 
the nature and distribution of Dark Matter, the mass of
neutrinos and the origin of the Cosmic Rays, to the existence of antimatter.
In addition the level and source composition of cosmologically important 
extragalactic diffuse radiation levels are probed and tests are
performed on non-Lorentz invariant interaction
and propagation terms predicted, e.g., in several ans\"atze for Quantum
Gravity theories.

Among the results and developments not covered here are the
latest developments in gravitational wave antennas (for a recent review see,
e.g., \cite{maggiore}), the results on atmospheric neutrinos
(see \cite{jung}), the measurements of the Cosmic Rays composition and
spectra around the 'knee' in the 
all-particle spectrum (see, e.g., \cite{utah}), or the results
from the first flight of the AMS detector \cite{hofer}.

The structure of this paper is as follows: section~\ref{sec:dm} reviews
the status of the Dark Matter search. The importance of Cosmic Ray
measurements at low and high energies are discussed in section~\ref{sec:cr}
followed by some of the latest developments in Gamma 
and Neutrino Astronomy in sections~\ref{sec:ga} and \ref{sec:na},
respectively. A brief outlook is given in section \ref{sec:ol}.

\section{The Dark Problems}
\label{sec:dm}

\begin{figure}[tbp]
\begin{center}
\vspace{5pt}
\psfig{figure=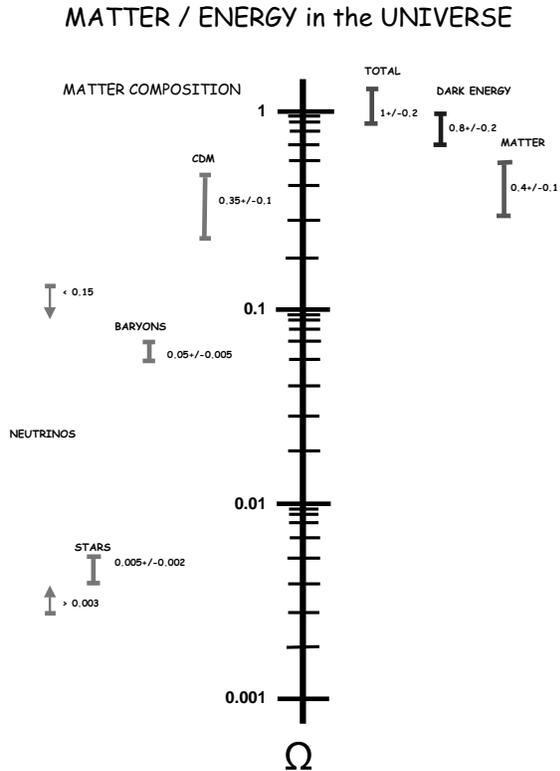,height=4.0in}
\end{center}
\caption{Current status of the observational and theoretical
census of matter and energy in the Universe. From \cite{turner}.}
\label{fig:dm1}
\end{figure}

Some recent reviews on cosmology, e.g., \cite{turner}, conclude that
for the first time now we have a \textit{complete} observational accounting
of matter and energy in the Universe consistent with $\Lambda$CDM (i.e.,
Cold Dark Matter plus non-zero $\Lambda$ term) inflationary cosmology. 
This, however, is not equivalent to a complete understanding
of these ingredients, and 
according to one summary shown in figure~\ref
{fig:dm1} the three \textit{Dark Problems} that we are facing today are:
Where are the dark baryons? What is the non-baryonic Dark Matter? What is
the Dark Energy recently detected in analyses of distant supernovae \cite
{sn1,sn2}? In the following the status of the 
questions concerning the amount and nature of Dark Matter (DM) 
will be addressed.

The evidence for DM is manifold and can roughly be divided into evidence on
galactic scales (i.e., $\mathcal{O}$(10) kpc) and on galaxy cluster scales
(i.e., $\mathcal{O}$(10) Mpc). 
Measurements of rotation curves of galaxies \cite{rubin},
especially of spiral galaxies including the
Milky Way, still give the most convincing evidence for the presence of a
galactic Dark Matter (DM) mass component which is about a factor 10 larger than the
observed luminous matter. Most of this mass has to be distributed in an extended
halo in an approximately spherical distribution with density falling as 1/$%
r^2$. 
Inspection of figure~\ref{fig:dm1} shows that this factor can in principle
be accommodated by baryons produced in Big Bang Nucleosynthesis (BBN). 
On scales of the extension of rich
galaxy clusters the second DM problem becomes visible. The main mass
fraction of rich clusters is in form of hot intracluster gas, the mass of
which can be inferred by measuring the X-ray flux or by mapping the
Sunyaev-Zel'dovich decrement caused by the scattering of cosmic
microwave background (CMB) photons on the
hot electrons. The total cluster mass is then derived by one or more of
three methods: from motion of the galaxies utilizing the virial theorem, by
assuming hydrostatic equilibrium for the hot gas, or, in some cases, by
gravitational lensing. From these data the mean matter density in the
Universe, $\Omega_M$, can be inferred. The obtained value of $\Omega_M$ =
0.4 $\pm$ 0.1 is consistent with values derived from other observables like
bulk matter flow or the peculiar motion of our Local Group of galaxies 
(e.g., \cite{turner}). Assuming BBN to be correct the main matter component in the 
Universe thus has to be \textit{non-baryonic}. 
Note that the validity of BBN is well tested by the 
measured $^4$He, D, $^3$He, and $^7$Li abundances (e.g., \cite{olive}).
In addition to the data there are solid theoretical
arguments for the existence of a large non-baryonic DM mass fraction in the
Universe. E.g., without large amounts of non-baryonic DM
there is no working theory of galaxy formation consistent with the small
level of fluctuations observed in the CMB. 

So far DM has only be discovered by its gravitational effects and the search for the
manifestation of DM in other interactions mainly is limited to searches for
galactic DM.
These searches
can either focus on the baryonic or a possible non-baryonic component.
Note that 
a debate on whether any non-baryonic component is present at galactic scales 
at all has been going on since the first solid evidence for galactic DM was obtained.

\subsection{Search for Baryonic Dark Matter}
\label{ssec:bdm}

Candidates for the baryonic component are low mass stars,
stellar remnants, Massive
Astrophysical Compact Halo Objects (MACHOs) such as brown dwarfs and
Jupiters with masses below $\sim$ 0.09 \mbox{M$_{\odot}$}, cold molecular
clouds, or very massive black holes. 
Of these candidates all but
the cold molecular clouds are in principle
detectable by {\it gravitational microlensing}.

\begin{figure}[tbp]
\begin{center}
\psfig{figure=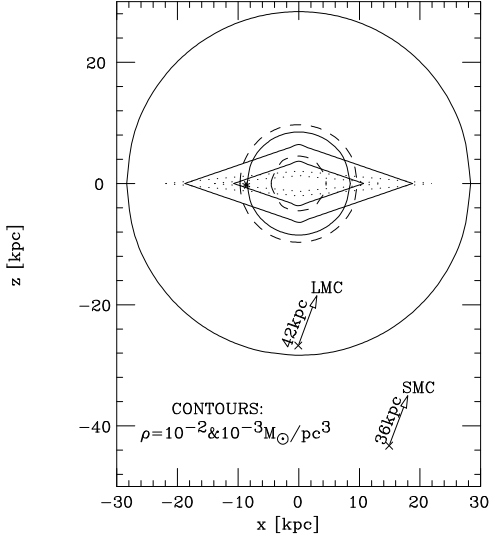,height=3.0in}
\end{center}
\caption{Schematic view of lensing populations discussed for the Galaxy: the
standard spherical halo (solid lines), a heavy spheroid (dashed lines), a
maximum thick disk (solid lines) and a dark thin disk (dotted lines). For
each population the density contours $\rho$ = 10$^{-3}$ and $\rho$ = 10$%
^{-2} $ \mbox{M$_{\odot}$}/pc$^3$ are shown. The locations of the sun and
the small (SMC) and large (LMC) Magellanic Clouds are indicated. From 
\protect\cite{roulet}.}
\label{fig:dm2}
\end{figure}

Gravitational microlensing (ML) as a means to search for MACHOs 
was proposed in a seminal paper by Pacy\'{n}ski \cite{paczynski} in 1986.
Only 13 years after this proposal ML is a thriving field and
can be considered a new field of \textit{galactic astronomy} (for a review see, e.g.,
\cite{roulet}).
Several different target directions have been monitored in
the search for the transient ML events characterized by the achromatic,
symmetric and unique amplification of a source star. The pioneering work was
performed by the EROS and MACHO collaborations who in the last years
monitored millions of stars in the small (SMC) and large (LMC) Magellanic
Clouds (see figure~\ref{fig:dm2}) and announced the first candidate events in 1993 
\cite{aubourg,alcock}. The total ML events statistics towards the SMC and
LMC currently is about 20 events (from EROS, MACHO, and OGLE). Other target
directions have been the galactic bulge (e.g., OGLE, OGLE-2, DUO, MOA, and
MACHO), galactic spiral arms (e.g., EROS), or towards the
Andromeda galaxy (AGAPE). In addition new collaborations (e.g., MACHO/GMAN,
MPS, or PLANET) aim at accurate photometry of ML events in order to 
search for planets in the source star systems. The ML event statistics
towards galactic targets is currently more than 400 and is utilized to study
details of galactic structure unobservable by other methods (see, e.g.,  \cite
{moniez}).

For the interpretation of ML events in terms of DM one has to take into
consideration that the velocity, distance and MACHO mass degeneracy in the
basics ML equations (see Eq.~(1) and (2)) does not allow to determine the
location of the lenses itself based on the magnification and duration of the
event alone. Only for special events like binary ML events discovered
towards the Magellanic clouds and long duration events showing parallax
effects due to the movement of the earth around the sun is it possible to
constrain the location of the lenses. E.g., for the binary-lens caustic
crossing event MACHO-98-SMC-1 which was monitored by 5 ML collaborations
after the MACHO alert, the combined analysis yields strong evidence for the
lens system to be located in the SMC itself \cite{afonso}.

In general, however, the basic ML equations only relate the velocity,
distance and mass of the lenses to give the lensing probability: 
\begin{equation}
\tau \propto \int \rho(x) x(1 -x) dx
\end{equation}
with the mass density distribution, $\rho (x)$, of microlensing matter needed
as an input according to the adopted \textit{isotropic halo model} (see
figure~\ref{fig:dm2}), and $x$ = $D_{lens}$/ $D_{source}$ with $D_{lens}$
and $D_{source}$ the distances of the lens and the source star,
respectively. The duration of the lensing event is given by: 
\begin{equation}
\Delta t \propto \sqrt{\frac{M x(1 - x)}{v^2_{\perp}}}
\end{equation}
with $M$ the mass and $v_{\perp}$ the transverse velocity of the lens. The
lensing probability, $\tau$, gives the probability for finding a source star within
the Einstein radius, $R_E$, of some MACHO and which thus is magnified by a
factor A = 1.34. The smallness of $\tau$ ($\mathcal{O}$ (10$^{-6}$)) makes it
necessary to simultaneously monitor millions of stars. 

\begin{figure}[tbp]
\begin{center}
\psfig{figure=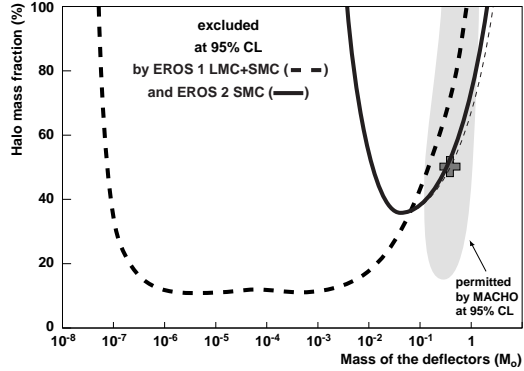,height=2.0in}
\end{center}
\caption{Exclusion diagram at 95\% CL assuming the standard spherical halo.
The dashed line is the limit from EROS 1 towards the LMC and SMC 
\protect\cite{renault}, the solid line is the limit from EROS 2 towards the
SMC \protect\cite{afonso2}. The 95\% CL allowed region from the MACHO
collaboration is shown as the shaded area \protect\cite{alcock2} with the
preferred value indicated by the cross. 
The thin dashed line corresponds to the limit obtained under the assumption
that no halo event has been observed. 
From \protect\cite{afonso2}.}
\label{fig:dm3}
\end{figure}

The combined limits and results on the galactic halo mass fraction in form
of MACHOs from the observations of the LMC and SMC by the EROS and MACHO
collaborations are shown in figure~\ref{fig:dm3} \cite{afonso2}. 
From the lack of short duration events it can be concluded that MACHOs in
the mass range 10$^{-7}$ \mbox{M$_{\odot}$} $%
\mbox{$\stackrel{\scriptstyle <}
{\scriptstyle \sim}$}$ $m$ $\lsim$ 10$^{-3}$ \mbox{M$_{\odot}$} \ make up
less than 25\% of the halo mass \textit{for most halo models}. On the
other hand, the MACHO collaboration derived a most probable MACHO mass range
of $m$ = 0.5$^{+0.3}_{-0.2}$ \mbox{M$_{\odot}$}\ from the observation of the
average long duration of 8 candidate ML events in the first 2.1 years of LMC
data \cite{alcock2}. In addition this event rate is compatible with about
50\% of the mass of the halo in the form of MACHOs of this mass. In figure~%
\ref{fig:dm3} the 95\% CL allowed halo mass fraction region as a function of 
$m_{\mathrm{MACHO}}$ derived from these data is also shown. Recently the
EROS collaboration published its results on the first 2 years of observation
towards the SMC \cite{afonso2}. Only one ML candidate event was observed. 
This low event
rate can be transformed into a limit on the fraction of the halo made out of
heavy MACHOs assuming a standard spherical halo model. The 95\% CL exclusion
limit is shown as the heavy solid line in figure~\ref{fig:dm3} and now
excludes a halo completely made out of heavy (i.e., $\sim$ 0.5 \Msol)
MACHOs. Note that these limits
are lower if some of the observed ML events are due to self-lensing. 
Self-lensing means that the lenses are not in the halo but in 
tidally elongated Magellanic clouds themselves
or in the (warped) disk of our own galaxy.
In addition the lensing events might be due to faint stars of
an yet undiscovered intervening dwarf galaxy. If the ML
events should be caused by one of the above effects MACHOs will be irrelevant for DM
whatever their astrophysical nature. In order to study this further and get
a better handle on the location of the lenses through the comparison of mass
moments extracted from the ML data with halo models (e.g., \cite{roulet})
a much larger event statistics towards the LMC and SMC and other
line-of-sights are necessary.

Independent constraints on the DM contribution of baryonic compact objects
come from astrophysics and cosmology which now rule out most of the compact
candidates for baryonic DM in the halo. Without discussing each limit in
detail the situation of protostellar objects, stellar objects and stellar
remnants as candidate objects for making up the halo mass can be summarized
as follows:

\begin{itemize}
\item  Jupiters (10$^{-7}$ $<$ M $<$ 10$^{-2}$ \mbox{M$_{\odot}$}) are ruled
out by ML data, i.e., halo mass fraction $<$ 25\%.

\item  faint hydrogen burning stars (M $>$ 0.09 \mbox{M$_{\odot}$}) and
young brown dwarfs (M $<$ 0.09 \mbox{M$_{\odot}$}) are ruled out on the
basis of HST and ISO data and contribute a halo mass fraction of $%
\mbox{$\stackrel{\scriptstyle <}
{\scriptstyle \sim}$}$ 0.01 \cite{gilmore}.

\item  old brown dwarfs (M $<$ 0.09 \mbox{M$_{\odot}$}) are ruled out by HST
and US Naval Observatory parallax data and star formation theory and
contribute a fraction $%
\mbox{$\stackrel{\scriptstyle <}
{\scriptstyle \sim}$}$ 0.03 \cite{freese}.

\item  white dwarfs (wd), neutron stars, and black hole stellar remnants are
ruled out by observations strongly constraining the number of possible
progenitor stars. The constraining data range from chemical abundances of C
and N measured in Ly$\alpha $ systems \cite{fields}, global D and $^{4}$He
abundances \cite{fields2}, to the stringent limit on the
cosmic infrared background flux inferred from the
observation of multi-TeV $\gamma $-rays
from the blazar Mrk 501 at a redshift of $z$ = 0.034 \cite{funk} .
All of these signal levels would have been raised by 
progenitor stars, yielding:

\begin{itemize}
\item  $\Omega _{wd}$$h$ $\leq $ 2$\times $10$^{-4}$ \cite{fields}

\item  $\Omega _{wd}$$h$ $<$ 0.003 \cite{fields2}

\item  $\Omega _{wd}$$h$ $\leq $ (1-3) $\times $10$^{-3}$ \cite{graff}
\end{itemize}
with the Hubble parameter H$_0$ = $h$ $\cdot$ 100 km s$^{-1}$ Mpc$^{-1}$.
\end{itemize}

In conclusion, the ML data in combination with current astrophysics seem to
rule out a completely baryonic halo, \textit{if} it is made out of
isotropically distributed compact massive objects. The nature of the
observed events towards the LMC and SMC remains unknown. One possible
explanation is a halo made out of {\it primordial} black holes with a mass of
about one solar mass which would explain the ML events and act as cold DM
otherwise. Other exotic candidates like mirror matter stars which would also
circumvent the astrophysical limits are discussed in the literature (e.g.,
\cite{volkas}). An astrophysics scenario which would still
allow for a purely baryonic halo is a halo made out of cold molecular clouds
(see, e.g., \cite{paolis}) without explaining the ML events. 
However, most of the data and theoretical models
currently point to a substantial fraction of the halo being made up by
non-baryonic DM.

\subsection{Search for Non-Baryonic Dark Matter}
\label{ssec:ndm}

Experiments have to answer the question of the nature of the non-baryonic DM
(NBDM). The search for observable signals resulting from interactions other
than gravity in general require detectors especially suited to the DM
candidate in question.

At the moment there is a surplus of matter (types) in cosmology and at
the same time important stable and massive particles predicted to exist
either within the Standard Model or in its plausible extensions have not yet
been discovered. The most promising approach to searches for NBDM therefore
lies in performing dedicated searches for selected Particle Physics
candidates. These have to fulfill the requirement to have decoupled from the primeval
plasma when their interaction rate became smaller than the cosmic expansion
rate and which therefore would be floating around the Universe today as 
\textit{relic particles}. According to their energies at the time of
decoupling one differentiates between relativistic (for masses less than 1
keV), i.e., hot, and non-relativistic, i.e., cold, relics. The best
candidates which could contribute significantly to
the mass-energy density in the Universe (i.e., $\Omega$ $\sim$ 1)
are the axion, the neutrino, the
lightest supersymmetric particle if R-parity is conserved, or magnetic
monopoles. Note that monopoles do not work well in numerical simulations of
structure formation and are thus regarded to be the least probable solution
to the NBDM problem by most cosmologists and we shall in the following not
discuss experimental searches for monopoles.

Of the remaining candidates the axion is a very good candidate but hard to
detect in laboratory experiments within the current mass constraints. It is
the Goldstone boson of the U(1) Peccei-Quinn symmetry \cite{peccei} as the
elegant solution to the strong CP problem of QCD. At the same time it would
be a good cold DM (CDM) candidate as, within the current laboratory and
cosmological limits, relic axions could make up a significant fraction of
the critical density. The sensitivity of experiments searching for light
axions via the Primakoff conversion of axions into microwave photons in
strong magnetic fields are now entering into the cosmological interesting
range \cite{hagemann}. See~\cite{axion} for a detailed review of the recent
development and the current limits which mainly stem from astrophysics and
cosmology

The next candidate, a light massive neutrino, would be a hot DM (HDM)
particle and its mass would be directly proportional to its contribution to $%
\Omega$. More precisely, the sum of the mass eigenstates which contribute to
the active flavour neutrinos is directly proportional to $\Omega_{\nu}$. In
order for the structure formation theories to work neutrinos (or other HDM)
cannot be responsible for the whole non-baryonic DM fraction but have to
give a contribution to $\Omega$ of $\lsim$0.15. 
This can be translated into $\sum$ $M_i$ $\lsim$6 eV 
(for H$_0$ = 65 km $\cdot$ s$^{-1}$ $\cdot$ Mpc$^{-1}$
and $i$ = 1, $\cdots$, 3) \cite{turner}. This number, derived
from a cosmological measurement, i.e., the fluctuation spectrum of the
cosmic microwave background, is compatible with (and competitive to) the
direct determination of the mass of the electron antineutrino in the
measurement of the endpoint of the $\beta$ spectrum in tritium decay. 
These experiments now start to consistently yield a positive
antineutrino mass in the fit, with the current value from the Mainz
experiment reported at this conference of $m_{\bar{\nu}}$ $\leq$ 2.8 eV
(95\% CL) \cite{weinheimer}. In spite of the large relic neutrino density
(54 cm$^{-3}$ per light flavour) the determination of their exact contribution
to $\Omega$  will most
probably come from this type of direct laboratory experiments, 2$\beta$
decay experiments (e.g., \cite{klapdor0}), or the forthcoming long-baseline
oscillation experiments (e.g., \cite{jung}). However, as the 2$\beta$
decay experiments only give a mass convolution involving unkown phases and
oscillation experiments only give mass differences, the improvement of the
cosmological limits will continue to be important.

The fundamental questions that are left open by the Standard Hot Big Bang
Cosmology point towards a grander theory. The best candidate to date is
Inflation plus cold Dark Matter leading to a flat and slowly moving
Universe. In addition Particle Physics models beyond the Standard Model
offer a variety of CDM candidate particles. As Superstring theories predict
a Supersymmetry sector at the TeV scale the favourite CDM candidate in
extensions of the Standard Model is the Lightest Supersymmetric Particle
(LSP), presumably the neutralino, in minimal supersymmetric models with
conserved R-parity. If these particles have been produced in the early
Universe and have decoupled when they became non-relativistic their density
would be inversely proportional to their interaction rate. The requirement
that the LSP contributes to $\Omega$ of $\mathcal{O}$(1) makes it weakly
interacting with a mass between $\sim$10 GeV and $\sim$500 GeV \cite
{jungman}, hence Weakly Interacting Massive Particle (WIMP). 

Assuming that WIMPs contribute a substantial fraction to the galactic halo
mass leads to a density at the location of the earth in the range of 0.3 -- 0.4
GeV cm$^{-3}$ in the
isothermal standard halo picture (depending slightly on the chosen halo parameters). 
WIMPs can be detected directly or
indirectly, where their elastic scattering on matter constitutes the direct
channel and the detection of annihilation products expected from
pairwise annihilation processes define the indirect channels.

The direct searches are based on the detection of nuclear recoils of $%
\mathcal{O}$(keV) as signals of the elastic scattering process within the
sensitive detector volume. The expected signal distribution is roughly
exponential with the mean depending on the mass of the WIMP. The expected
signals rates are of the order of 1 event per day and kg of detector mass.
In the scattering only a small fraction of the delivered energy, however,
goes into the ionization channel. Most of the energy is released as
heat and cryogenic devices are therefore intensely studied for their
suitability. The main background is due to electron recoils from Compton
scattering of background $\gamma$-rays. Due to the motion of
the earth through the WIMP halo an annual modulation of the direct scattering signal with
an amplitude of the order of a few \% is expected.

\begin{figure}[tbp]
\begin{center}
\psfig{figure=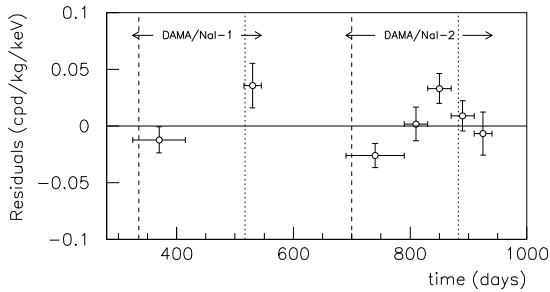,height=1.9in}
\end{center}
\par
\vspace{-0.5cm}
\caption{Residual count rate versus time elapsed since January 1 of the first
year of DAMA data taking. The expected modulation is a cosine function with
minimum at December 2 and the maximum at June 2. From \protect\cite
{bernabei3}.}
\label{fig:dm5}
\end{figure}

\begin{figure}[tbp]
\begin{center}
\psfig{figure=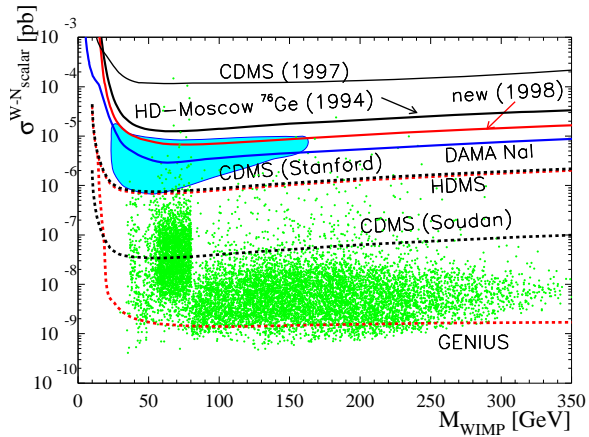,height=2.5in}
\end{center}
\caption{WIMP-nucleon cross section limits for scalar interactions as a
function of WIMP mass. The shaded area indicates the DAMA 90\% CL
evidence contour. The solid lines indicate excluded regions. The broken
lines indicate the expected sensitivity for the ongoing CDMS experiment at
different sites \protect\cite{akerib} and the proposed GENIUS experiment 
\protect\cite{klapdor}. See \protect\cite{baudis} for full references. From
\protect\cite{baudis}.}
\label{fig:dm6}
\end{figure}

Detectors with a low inherent radioactive background like Germanium, 
NaI, or cryogenic detectors are mostly used to
search for WIMP scattering signals (e.g., \cite{ws} for a review). 
The DAMA
collaboration presented a hint for an annual modulation signal in 1997 
\cite{bernabei} and recently announced  $\sim$3$\sigma $ evidence for annual
modulation after quadrupling the analysed statistics \cite{bernabei2}. The
residual detector count rate versus time is shown in figure~\ref{fig:dm5}.
The 90\% CL allowed region of scalar
WIMP-nucleon scattering cross section versus WIMP mass 
only barely enters into the region expected in constrained minimal
supersymmetric models (CMSSM) (see figure~\ref{fig:dm6}). 
The constrained or 'phenomenological' MSSM is the simplest
supersymmetric extension of the Standard Model with the
number of free parameters of the most general R-parity conserving model
of 124 reduced to 7 real parameters.
Two other NaI experiments, the UK Dark Matter Collaboration (UKDMC) \cite
{smith} which is taking data since 1996 in the UK Boulby Mine and a French
experiment at Modane \cite{gerbier} studying NaI(Tl) crystals for
their suitability as DM
detectors, use the PMT pulse shapes to identify and distinguish against $%
\alpha $, $\beta $, and $\gamma $ backgrounds. Both experiments find a not yet
understood population of short pulses not consistent with background
expectation according to elaborate background studies (e.g., \cite{smith2}). 
These data are also not fully compatible to the WIMP
signal expectation and are thus not interpreted in terms of a WIMP signal. 
Due to the poor intrinsic background rejection capability of
NaI detectors rather some hidden systematics is suspected.
In order to test the DAMA
evidence in future searches the active identification of the background will
be of foremost importance (e.g., with liquid Xenon or cryogenic
devices). Note that, e.g., the current UKDMC sensitivity is comparable to DAMA. 
Some of the existing limits on the scalar WIMP-nucleon cross
section versus WIMP mass together with some expected
sensitivities for some of the planned experiments are shown in figure~\ref{fig:dm6}. 

\begin{figure}[htbp]
\begin{center}
\psfig{figure=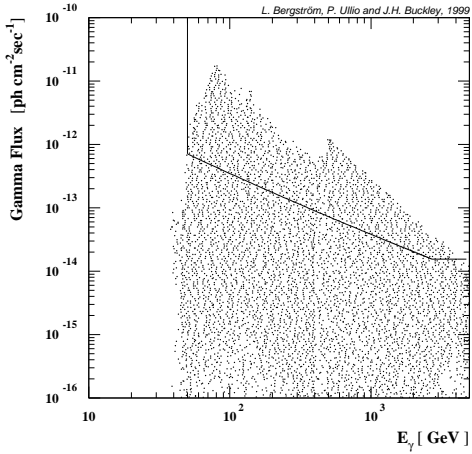,height=2.6in}
\end{center}
\vspace{-10pt}
\caption{Flux sensitivity (5 $\sigma$ in 50 h observation time) 
of the upcoming MAGIC Telescope (full line) for a $\gamma$-ray source
at the position of the GC compared to flux calculations
within CMSSMs. Each point represents
the prediction for one possible CMSSM model representation. 
All points satisfy 0.025 $\leq$ $\Omega_{\chi} h^2$ $\leq$ 1.
From \cite{bergstroem3}.}
\label{fig:dm7}
\end{figure}

\begin{figure}[tbp]
\begin{center}
\psfig{figure=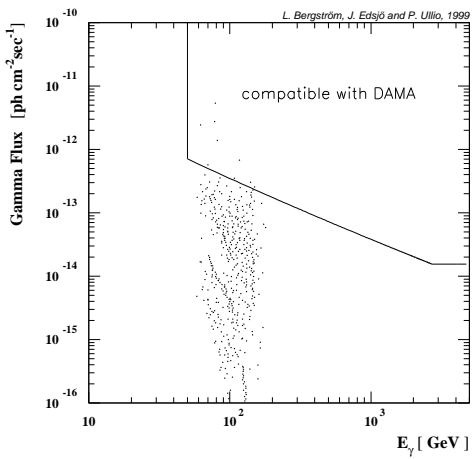,height=2.6in}
\end{center}
\vspace{-15pt}
\caption{As for figure~\ref{fig:dm7} but only the flux predictions
for those CMSSM models are shown which predict a scalar elastic scattering cross
section compatible with the DAMA evidence contour. From \cite{ullio3}.}
\label{fig:dm8}
\end{figure}

Indirect searches for WIMP DM are based on the detection of the products
of pairwise WIMP annihilation in the galactic halo or in the
centre of earth or sun. This process will occur if the WIMP is the
supersymmetric neutralino, $\chi$, i.e., a Majorana
fermion. One of its characteristics is the production of equal amounts of 
matter and antimatter and all stable annihilation products,
e.g., $p$, $\bar{p}$, $e^-$, $e^+$, $\bar{D}$, $\nu$,
may serve as signature. In addition $\gamma$s can be produced in loop-induced
annihilation reactions \cite{bergstroem, ullio}. 
Whereas protons and electrons are too abundant in
the normal Cosmic Rays, antimatter, which does not exist in sizable
quantities in the observable Universe, may herald the annihilation of 
neutralinos in the galactic halo. The expected annihilation rate depends
on the WIMP density distribution in the halo. According to the halo models
shown in figure~\ref{fig:dm2} and according to results of N-body
simulations of the formation of dark halos
a considerable enhancement of the DM density near the Galactic centre (GC)
is expected. For the $\chi$ $\chi$ $\rightarrow$
$\gamma$ $X_0$ channel this will lead to a much enhanced signal from the direction of the GC 
where $X_0$ is a neutral particle of mass $m_X$ and the $\gamma$'s will be
nearly monochromatic with an energy
\[
E_{\gamma} = M_{\chi} - \frac{m_X^2}{4 M_{\chi}}.
\]
Due to the non-relativistic velocities
of the WIMPs this channel thus is characterized by a huge signal peak-to-width ratio
with no known astrophysical background source. The proposal to use ground-based
Imaging Air Cherenkov Telescopes (IACTs) to search for this signal dates back
to 1992 \cite{urban} when it was realized that the only average energy resolution
of the order of 10-20\% is offset by the huge collection areas of the order
of 10$^4$-10$^6$ m$^2$. According to recent calculations
\cite{bergstroem3} the upcoming generation of IACTs (see section~4)
will provide enough sensitivity to probe a significant part
of the parameter space of the CMSSM.
In figure~\ref{fig:dm7} an estimate of the flux sensitivity of the
MAGIC Telescope currently under construction on the Canary Island
La Palma \cite{lorenz} is shown in comparison to the flux predictions for 
the channel  $\chi \chi$ $\rightarrow$ $\gamma \gamma$ for a large number
of CMSSM models \cite{bergstroem3}. 
In addition figure~\ref{fig:dm8} shows the expected coverage
for the CMSSM parameter space
covering the DAMA evidence region illustrating the somewhat orthogonal
sensitivity to the supersymmetric parameter space in terms of direct and
annihilation cross section coverage for the same mass range.

A final remark on the DM problem: recent developments in experimental and
phenomenological Particle Physics point towards the fact that
the existence of the observed baryon asymmetry itself is a signature of
physics beyond the Standard Model.
Csikor et al. showed in a 4-dim.\,lattice calculation \cite{csikor, csikor2} that
there is no first order electroweak phase transition in the Standard Model if m$_{%
\mathrm{Higgs}}$ $\mbox{$\stackrel{\scriptstyle >}
{\scriptstyle \sim}$}$ 75 GeV.
The critical point in the
phase diagram of the electroweak phase
transition is thus below the current lower Higgs mass limit from LEP II 
\cite{gross}. Therefore one of Sacharovs
conditions for producing the observed baryon asymmetry (baryon number violation,
CP violation, and out-of-equilibrium dynamics as in a first order phase
transition) is not fulfilled for the electroweak symmetry breaking phase.
Another mechanism which may be responsible for the baryon asymmetry and which
has received growing attention lately due to the experimental evidence
for a non-zero neutrino mass, is the leptogenesis mechanism \cite{fukugita}.
In this model the cosmological baryon asymmetry is generated from a primordial
lepton asymmetry which was produced by the out-of-equilibrium decay
of heavy Majorana neutrinos.
In a leptogenesis scenario Bolz,
Pl\"umacher and Buchm\"{u}ller \cite{bolz} showed that 
without encountering the 'gravitino problem'
the non-baryonic DM might be present in the form of gra\-vitinos (as the lightest
supersymmetric particle) with masses in the range between
10 GeV and 100 GeV. This would mean that the nonbaryonic DM
would be present as GIMPs, i.e., Gravitationally Interacting Massive Particles, 
in which case all laboratory searches would be negative and only astronomy, 
e.g., through weak gravitational lensing and ML, could provide more data on the DM.

\section{Cosmic Rays and Antimatter}
\label{sec:cr}

In the last years advancements in detector technology have led
to an intensified study of many aspects of Cosmic Ray physics.
With the upcoming Cosmic Ray (CR) experiments
like, e.g., AUGER, AMS, and PAMELA, and the upcoming $\gamma$-ray detectors,
e.g., CANGAROO II, GLAST, HESS, MAGIC, and VERITAS, it is safe
to predict that this trend will continue (see also section~4). 
As the 'beam' of CRs is either background or signal
the understanding of the CR spectrum and composition
is of utmost importance for many 
current and upcoming astroparticle physics experiments.

Here we only report on some selected
results and a few of the upcoming experiments focusing on the
low energy and highest energy part of the measured 
CR energy spectrum.
In the low energy domain the interest in CR measurements, 
beyond the understanding of the sources and propagation of
the CRs themselves, 
is focussed on the derivation of the neutrino oscillation parameters,
the measurement of the antiproton and positron fluxes in the search
for neutralinos and primordial black holes, and the search for antimatter.

\begin{figure}[tbp]
\vspace{-20pt}
\begin{center}
\psfig{figure=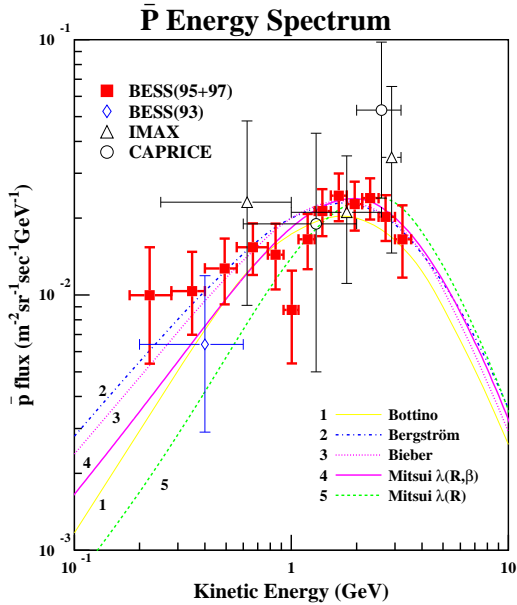,height=4.0in}
\end{center}
\vspace{-30pt}
\caption{BESS 1995 and 1997 (solar minimum) antiproton fluxes measured at the 
top of the atmosphere. Also shown are some of the previously existing data.
The error bars correspond to the quadratic sums of the statistical and systematic
errors. The curves represent some recent calculations of the expected
spectra from CR interactions with interstellar matter for the
solar minimum period. From \cite{yoshimura2}.}
\label{fig:cr3}
\end{figure}

A large fraction of past and present data on the absolute CR flux come
from balloon-borne experiments. A representative detector flown
in this type of experiment is the Japanese BESS spectrometer.
%shown in figure~\ref{fig:cr2} in its 1997 configuration. 
The main parameters and components of this detector are the 1 Tesla magnetic field
produced by a thin (4 g/cm$^2$) superconducting coil filling a tracking volume
equipped with drift chambers providing up to 28 hits per track with
an acceptance of 0.3 m$^2$ sr. In addition two hodoscopes provide
d$E$/d$x$ and time-of-flight measurements. In the 1997 flight data at the
top-of-the-atmosphere (mean residual air mass 5.3 g/cm$^2$)
were collected for 57,000 s. The combination of the measured $\bar{p}$ fluxes
as a function of kinetic energy from the 1995 and 1997 campaigns
are shown in figure~\ref{fig:cr3} \cite{yoshimura2}. 

Antiprotons should be produced as secondaries
in interactions of the galactic CR protons with the interstellar medium.
Their kinetic energy spectrum is expected to show a
characteristic peak around 2 GeV, with sharp decreases in flux towards
smaller and larger energies. Other possible sources of $\bar{p}$
are the annihilation of neutralinos at the GC or the evaporation
of primordial black holes (PBH).
The latter possibility received growing attention in the recent past
following the detection of antiprotons with kinetic energies below 0.5 GeV
by BESS in 1993 \cite{yoshimura}. 
Today black holes are only expected to be formed in stellar collapses
of stars with several solar masses. The uniformity of space-time precludes
collapses of matter to black holes with masses below this limit.
In the early Universe, however, virulent conditions may have led
to the formation of PBH with arbitrarily small masses \cite{novikov}, e.g.,
by the collapse of large density perturbations
\cite{zeldovich, hawking}. Data constraining PBH abundances
will thus yield constraints on the density fluctuation spectrum
in the early Universe, an important ingredient of structure
formation theories. 
Quantum effects lead to the evaporation of these PBH by particle emission \cite{hawking2}.
The emission spectrum is similar to a black body with finite size
and a temperature 
\[
T_{PBH} = \frac{\hbar c^3}{8 \pi G M_{PBH}} = 1.06 \left( \frac{10^{13}\mathrm{g}}{M_{PBH}}
\right) \mathrm{GeV}
\]
where $M_{PBH}$ is the mass of the PBH in grams. For $T_{PBH}$ above
the QCD-scale of $\sim$ 300 MeV the evaporation process will result in
relativistic quarks and gluons which may produce antiprotons
during hadronisation. The expectation for the flux of
antiprotons from this process is a spectrum increasing towards
lower kinetic energies down to $\sim$0.2 GeV \cite{macgibbon}.
Similarily the spectrum of $\bar{p}$ expected from the annihilation
of neutralinos at the GC is characterized by a significant
flux below 1 GeV \cite{bergstroem4}. 
The uncertainty of the expected secondary flux
due to the uncertainty of CR propagation in the Galaxy, however, is still
of the same order of magnitude as
the 'signal' expected from neutralino annihilation.

This propagation uncertainty is illustrated in figure~\ref{fig:cr3}. The
data points show that BESS now has 
convincingly measured the predicted peak in the kinetic energy 
distribution at $\sim$2 GeV. The location and height of this peak can
be calculated by using the measured $p$ and He CR spectra 
together with the cross sections for $\bar{p}$ production
measured at accelerators. The different propagation
model calculations shown in figure~\ref{fig:cr3} differ mainly in propagation
parameters.
Although the data below 2 GeV now are much better from the statistical
point-of-view, the uncertainty in the propagation of secondary $\bar{p}$s
still precludes any firm conclusion on additional 'direct' sources of $\bar{p}$s.
Future measurements at times with higher solar activity will 
be performed, e.g., by BESS, in order to study the secondary component in more detail.
At times of high solar activity the 'direct' $\bar{p}$ component should
be virtually absent.
In addition extending the measurements
to larger energies (i.e., $E_{kin}$ $\gg$ 2 GeV) 
will also yield constraints on the propagation models. 
Here the current experiments
did not have enough sensitivity and exposure to limit the propagation parameters.
Upcoming CR experiments like, e.g., the PAMELA experiment, will here yield
much improved data.
PAMELA \cite{adriani} is a satellite-borne magnet spectrometer 
currently being built by the WiZard collaboration together with
Russia. It is planned to be launched from Baikonur in 2002.
The scientific objectives of the PAMELA mission are to measure the spectra of $\bar{p}$,
e$^+$, and nuclei in a wide range of energies, to search for primordial
antimatter and to study the CR fluxes over half a solar cycle. PAMELA will be able to
measure particles with magnetic rigidities (momentum/charge) up to a few hundred GV/c.

Contrary to the antiproton signal the positron signal from neutralino
annihilation at the GC is expected to show up at large kinetic energies,
i.e., well above the geomagnetic cutoff. The expectations for 
improved data in this channel from the upcoming experiments are also high.
Similar to the antiproton channel the current data base on CR positrons does not yet
allow any conclusion on additional sources besides CR interactions
in the interstellar medium. 

Another important aspect of CR physics is the search for primordial antimatter.
This search mostly focuses on antihelium nuclei
and currently the upper limit on the CR anti-helium/helium ratio
provides the best evidence for the Galaxy and the nearby Universe
being made up solely of matter.
The combination of the BESS collaboration data taken during 5 campaigns
between 1993 and 1998 have led to an improvement of the upper limit
on the ratio of the abundances of antihelium to helium
with magnetic rigidities between 1 GV/c and 16 GV/c by about a factor
of 90 \cite{nozaki}. 
The upper limit of 10$^{-6}$ quoted by BESS is compatible
with the limit presented by AMS \cite{hofer} and any further substantial
improvement will need long duration measurements, e.g., by PAMELA
or AMS.

The absolute fluxes of proton, helium and atmospheric muons
are important for the derivation of the neutrino
oscillation parameters since the atmospheric 
neutrino flux is proportional to the normalization
of the dominating CR proton and helium fluxes. 
In addition there other applications where
the absolute measurement of the proton spectrum is 
important, e.g., in predictions of the secondary
antiproton and positron fluxes needed to constrain
exotic production channels and
the above-mentioned propagation models.

\begin{figure}[tbp]
\begin{center}
\psfig{figure=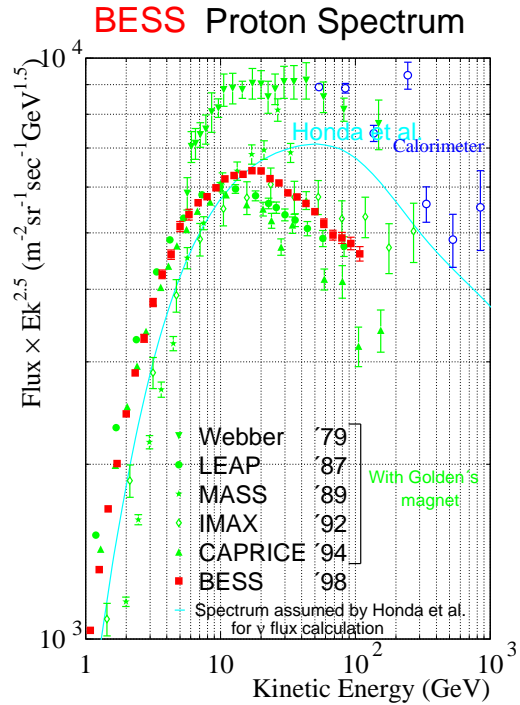,height=4.0in}
\end{center}
\vspace{-20pt}
\caption{The absolute differential CR proton spectrum as measured by the BESS collaboration
in 1998. Also shown are some previous measurements, and, as the full line,
the spectrum assumed in the Honda {\it et al.}\, \cite{honda} calculation
of the atmospheric neutrino fluxes. From \cite{sanuki}.}
\label{fig:cr7}
\end{figure}

Figure~\ref{fig:cr7} shows the most recent 
measurement of the CR differential proton spectrum
in the kinetic energy range between 1 GeV and $\sim$100 GeV 
by BESS together with data taken by previous experiments \cite{sanuki}. Shown as the full line
is the proton spectrum assumed in the Honda {\it et al.}\, Monte Carlo calculation \cite{honda}
used in the analysis of the SuperKamiokande data on atmospheric neutrinos \cite{fukuda}. 
The different data sets show discrepancies in the measured absolute fluxes 
as large as a factor of 2 at $\sim$50 GeV.
The absolute differential proton flux 
was also measured with the MASS balloon detector in 1991 but published only 
recently \cite{bellotti}. In addition the differential helium flux at the top of the 
atmosphere and the differential
muon fluxes in the atmosphere as a
function of atmospheric depth were measured.
The proton and helium fluxes above 10 GeV were found to be compatible
with the LEAP '87 data \cite{seo} 
as shown for the proton case in figure~\ref{fig:cr7}
corroborating a lower normalization of the absolute proton
and helium fluxes as compared to previous
compilations of data, e.g., \cite{papini}. 

\begin{figure}[tbp]
\begin{center}
\psfig{figure=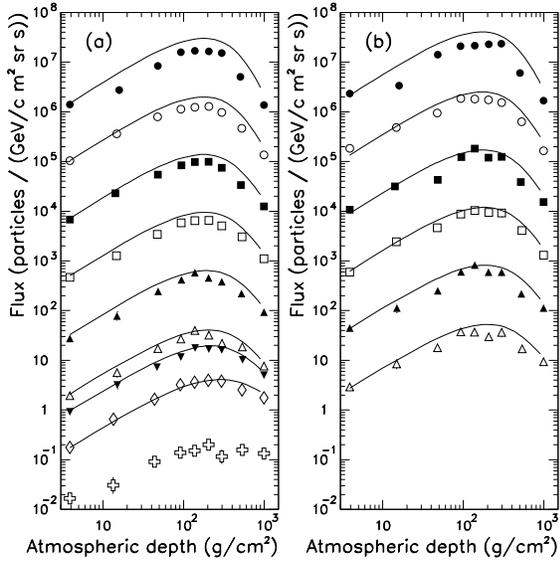,height=3.0in}
\end{center}
\caption{Differential (a) $\mu^-$ and (b) $\mu^+$ fluxes as a function
of atmospheric depth. From top to bottom the momentum ranges (in GeV/c) are:
0.3-0.53 (scaled by 10$^5$), 0.53-0.75 (10$^4$), 0.75-0.97 (10$^3$),
0.97-1.23 (10$^2$), 1.23-1.55 (10), 1.55-2 (1), 2-3.2 (1), 3.2-8 (1),
and 8-40 (1). The $\mu^+$ data are shown up to 2 GeV/c. The solid lines
represent the Bartol Monte Carlo calculation results\cite{bartol}. From \cite{boezio}.}
\label{fig:cr10}
\end{figure}

The CAPRICE94 balloon experiment measured 
the atmospheric muon flux
as a function of atmospheric depth \cite{boezio} and compared 
the data to
calculations using the Bartol Monte Carlo \cite{bartol} (figure~\ref{fig:cr10}).
The disagreement between data and Monte Carlo prediction
increases with atmospheric depth from about 1.1 $\pm$ 0.1 (stat.) $\pm$ 0.1 (syst.) 
for the ratio of simulation result to the measured $\mu^-$ flux
at the top of the atmosphere up to  1.8 $\pm$ 0.1 $\pm$ 0.1 at the maximum
of the muon flux at a depth of around 200 g/cm$^2$. As the atmospheric muon flux
is closely related to the flux of atmospheric neutrinos \cite{perkins} these
data will have to be used to improve the Monte Carlo generators.
Gaisser investigated the influence of a change in slope and normalization
of the input proton spectrum on the atmospheric
neutrino result \cite{gaisser}. He concluded that for
a change in the flux normalization above 10 GeV from the
flux assumed by Honda {\it et al.}\,\cite{honda} in the direction of the new
BESS result the observed excess of electron neutrinos would be
increased by a similar amount and the deficit of muon neutrinos
correspondingly reduced. Taking these new data into consideration in
the fit of the SuperKamiokande data 
will very likely yield a shift in the fit parameters.
Such a shift could be of great importance for the upcoming long-baseline
neutrino experiments and illustrates the strengthening of the interconnection
of Particle Physics and Astroparticle Physics in the search for New Physics.

\begin{figure}[tbp]
\begin{center}
\psfig{figure=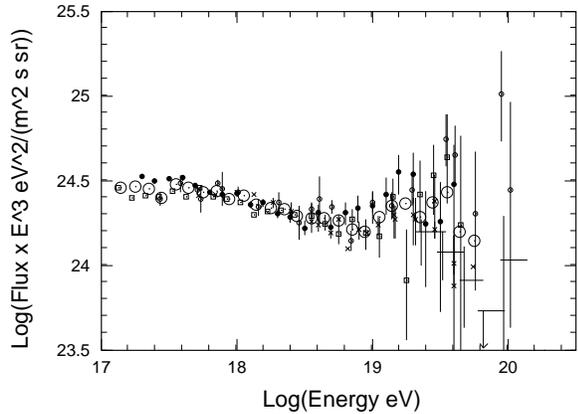,height=2.3in}
\end{center}
\caption{The top end of the CR energy spectrum as measured
by AGASA, Fly's Eye, Havarah Park, and Yakutsk.
The consistency of the spectrum is illustrated
by the fact that only a shift in the absolute energy scale of $\sim$15\% 
was necessary in order to align the spectra.
Note that the measured flux values were multiplied
by $E^3$ to enhance the 'ankle' feature at around 10$^{19}$ eV.  From \cite{yoshida}.}
\label{fig:cr11}
\end{figure}

At the uppermost end of the CR energy spectrum in the past 40 years
a number of experiments have been collecting events
with energies larger than 10$^{18}$ eV.
The galactic magnetic field (B $\sim$ 3 $\mu$G) cannot contain
CR protons with momenta larger than a few times 10$^{18}$ eV/c.
CRs with momenta above this value are thus very likely of extragalactic
origin.
Figure~\ref{fig:cr11} shows a compilation of the
all-particle flux above 10$^{17}$ eV as
measured by the four experiments AGASA, Fly's Eye, Havarah Park, and Yakutsk
until 1995 \cite{yoshida}.
The highest recorded CR energy to date is 3 $\times$ 10$^{20}$ eV \cite{flyeseye}.
Of the four experiments which contributed to figure~\ref{fig:cr11}
the AGASA (Akeno) group has reported new results this year.
The Akeno 20 km$^2$ air shower array was operated in Japan from 1984 to 1990 
and became part of the still operational $\sim$100 km$^2$ 
Akeno Giant Air Shower Array (AGASA)
in 1990.
The detection of an anisotropy in the arrival direction of CRs with energies 
around 10$^{18}$ eV (0.8 - 2 $\times$ 10$^{18}$ eV)
was reported by AGASA in 1998 \cite{hayashida} and 
updated recently \cite{hayashida2}.
The amplitude of the anisotropy in the first harmonic analysis was found to be 4\%
and could be identified in a 2-dimensional map as stemming from
4.5 $\sigma$ and 3.9 $\sigma$ excesses from the direction
of the Galactic Centre and the Cygnus region, respectively. 
This is a spectacular result as it is clear evidence for the existence 
of galactic CRs up to this very high energy. 
Two possible explanations for the observed anisotropy were put forward.
One is connected to the propagation of CR protons expected
in direction of the nearby spiral arm. The other is that the
anisotropy might be due to neutron primaries.
At energies of 10$^{18}$ eV neutrons have decay lengths
of $\sim$10 kpc and could thus propagate linearly
from the GC and Cygnus regions to the position of the earth without decaying.
More data are needed to distinguish between or rule out the models.
At larger energies the AGASA collaboration did not detect any significant
large-scale anisotropy with respect to the galactic or supergalactic plane \cite{takeda}.
The analysed event statistics was 581, 47, and 7 events with energies
larger than 10$^{19}$ eV, 4 $\times$ 10$^{19}$ eV, 
and 10$^{20}$ eV, respectively.

The size of the detectors and the time scale involved in collecting the 
data reported on above implies that only a new and bold ansatz will allow
to improve the event statistics in a significant way.
Especially at the uppermost end of the CR energy spectrum 
where 'New Physics' might be required to explain the data,
at least an order of magnitude improvement in statistics is called for.
Currently the world statistics of events with energies above 4 $\times$ 10$^{19}$ eV
is about 200 events corresponding to a total
integrated exposure of $\sim$1500 km$^2$$\cdot$yr$\cdot$sr. 
This energy is called the Greisen-Zatsepin-Kuzmin (GZK) cutoff energy and corresponds
to the proton energy for which the centre-of-mass energy in a collision with a cosmic 
microwave background (CMB) photon 
crosses the pion photoproduction threshold, i.e., 
\[
p + \gamma \rightarrow p + \pi^0 \qquad \mathrm{or} \qquad p + \gamma \rightarrow n + \pi^+.
\]
The survival probability of ultrahigh energy CR (UHECR) protons of
energies 8 $\times$ 10$^{18}$ eV, 10$^{20}$ eV, and 3 $\times$ 10$^{20}$ eV
after having traversed a distance of 20 Mpc in the CMB are 0.70, 0.55, and 0.12, respectively.
Sources of the UHECRs thus have to be within
the GZK volume in our cosmic vicinity, i.e., closer than 30 to 50 Mpc.
The possible sources discussed in the literature
range from nearby proton accelerators, i.e., powerful Active Galactic Nuclei (AGN)
\cite{biermann}, Gamma-Ray Bursts (GRBs) \cite{waxman, vietri},
the decay of superheavy relic particles \cite{hill},
a new stable supersymmetric hadron with a mass of a few GeV, e.g.,
the $S^0$ as a $uds$-gluino bound state which would have a much longer
pathlength in the CMB than ordinary hadrons \cite{chung},
to extremely high energy neutrinos which would annihilate
with relic neutrinos to produce hadronic jets \cite{weiler}.
For all these types of sources the EHECRs should point back (within $\sim$ 5$^{\circ}$)
to their source(s) {\it if} the extragalactic magnetic fields are weaker
than $\sim$ 0.1 $\mu$G. Should these fields be stronger the only sources
from above for which the directional information is retained are
the new stable supersymmetric hadron and the extremely high energy neutrinos
interacting in the vicinity of the earth.
As recent determinations of the
magnetic field strength within the local (Virgo) supercluster indeed
point towards a local field strength of $\sim$ 0.5 $\mu$G (e.g., \cite{kronberg})
'conventional' local sources like AGN or GRBs 
might be 'hidden' due to large magnetic
deflections already over distances of the order of Mpc \cite{farrar}.

The experimental investigation of UHECRs will be lifted to a new 
level with the upcoming Pierre Auger
experiment \cite{boratov}. In March of 1999 the ground breaking ceremony for the
southern detector took place in the
Pampa Amarilla in Argentina. The detector will consist
of $\sim$1600 water Cherenkov tanks spread over $\sim$3000 km$^2$
with a 1.5 km grid spacing. This air shower detector will be overlooked by
three fluorescence detectors situated within the array.
The acceptance of the detector above 5 $\times$ 10$^{19}$ eV will be $\sim$14,000
km$^2$$\cdot$yr$\cdot$sr. Full operation of the detector is planned for
the beginning of the year 2003.

\section{Gamma Astronomy}
\label{sec:ga}

Gamma astronomy is the most recent addition to the spectrum of
20th century astronomy. This concluded the opening of the
{\it complete} electromagnetic
spectrum to astronomical observations.
Currently $\gamma$-ray astronomy is performed by
space-borne and ground-based $\gamma$-ray detectors in non-overlapping energy domains.
The space-borne detectors measure in the energy domain up to $\sim$10
GeV whereas the ground-based detectors are limited
to photon energies {\it above} $\sim$200 GeV. The remaining gap
in the electromagnetic spectrum will be closed by
the upcoming detectors from both sides, i.e., by the ground-based
and space-borne detectors enlarging their energy coverage. 
The number of $\gamma$-ray sources 
detected so far is about 270 in the energy domain below 10 GeV \cite{macomb}
and 13 sources (with a varying degree of certainty) with E$_{\gamma}$ $\gsim$ 200 GeV
\cite{weekes}.
Note that about 2/3 of the $\gamma$-ray sources
detected below 10 GeV have not yet been identified with known
astronomical objects. This class of 'EGRET Unidentified Sources'
thus is one of the main targets for upcoming $\gamma$-ray experiments.

Here we only report on some recent results from ground-based $\gamma$-ray
astronomy, i.e., results obtained with Imaging Air Cherenkov Telescopes (IACTs).
The IACT technique is based on the exploitation of the inherent
differences of $\gamma$ and (CR) hadron air showers developing in the
atmosphere.
As a function of energy a growing number of secondary particles 
in the air showers have energies in excess of the threshold
energy of E$_{thres}$ $\sim$ 21 MeV (at sea level) above which
Cherenkov radiation is produced by the passage of charged
particles through air. The emission of the Cherenkov light is
concentrated into a small angular region in the forward 
direction and each shower typically illuminates an area
of $\gsim$ 5 $\cdot$10$^4$ m$^2$ on the ground. Positioning the
IACT anywhere in this area will result in a detection of
the incident particle. The different images
of hadrons and photons allow background suppression
factors of $\sim$600 with stand-alone telescopes.
The suppression factor is larger by
about an order of magnitude when one points several 
(3 to 5) telescopes at the
same source in the stereoscopic observation mode
pioneered by the HEGRA collaboration at the Canary Island La Palma \cite{daum}.
Note that due to the indirect detection of the $\gamma$-rays interferometry,
however, is not possible.
The angular resolution of IACTs for individual events
is about 0.1 to 0.2$^{\circ}$
and the energy resolution is between 10 and 40\%.
As for the background suppression, using several detectors
in stereoscopic mode to measure the shower parameters,
at current energies (E $\gsim$ 200 GeV)
leads to improvements in both parameters \cite{daum}. 
The large effective area gained by the indirect detection is 
one of the major advantages of the IACT technique as it allows
to detect very low $\gamma$-ray fluxes. The major drawback
of this technique, on the other hand, is the low duty cycle.
Most IACTs are only operated during dark and moonless
nights resulting in a duty cycle of about 10\%. 

Due to the restricted field-of-views 
of the instruments (a few degrees times a few degrees) 
the observations can only be performed
in the pointing mode. In this mode the telescopes are oriented at possible
$\gamma$-ray sources with guidance from other wavelengths together with
model predictions for the very high energy (VHE) $\gamma$-ray channel.
One such model prediction concerns 
the important question of the origin of the CRs. In most models
Supernova Remnants (SNRs) are the favoured
sites of CR acceleration. 
As the CR nuclei themselves do not
retain directional information in the galactic
magnetic field a neutral particle signal
must be searched for in order to identify the CR sources.
Whenever hadrons are accelerated to very high energies
pions should be produced as secondaries.
However, SNRs are known sources VHE $\gamma$-rays and the main
$\gamma$-ray component is
due to inverse Compton scattering of ultrarelativistic electrons
on low energy photons (e.g., CMB photons).
In order to identify the sources of the CRs
one thus aims at identifying an
{\it additional VHE $\gamma$-ray component}
stemming from $\pi^0$ decay.
SNRs thus have been intensively studied by
the IACTs in the past.
Although six SNRs have been observed above $\sim$200~GeV \cite{weekes}
(Crab nebula, Vela, PSR1706-44, SN1006, RJX1713.7-3946, and (possibly) CasA), 
this additional component has not been clearly identified, yet.
More sensitive measurements at lower energies will be
of great importance in identifying the spectral component(s)
observed in the six sources above and to discover $\gamma$-ray
emission in more SNRs. With low energy thresholds, e.g., 
of the upcoming MAGIC Telescope (initially 25 GeV),
it may be possible to observe a {\it two component $\gamma$-ray
spectrum}, which should then allow to decouple the predicted leptonic and
hadronic components in SNR shells.
In summary, no clear evidence for proton acceleration in SNRs was found up to now
and the sources of the galactic CRs are not identified yet.

The most spectacular discoveries in ground-based $\gamma$-ray astronomy
are related to the {\it extragalactic} sources of 
VHE $\gamma$-rays. The first such source discovered in 1992
by the Whipple collaboration \cite{punch} was the closest BL Lacertae (BL Lac) type
object Markarian 421 (Mrk 421) at a redshift $z$ = 0.031.
This AGN at the same time is one of the weakest $\gamma$-ray source detected 
by the EGRET experiment at energies below $\sim$10 GeV \cite{macomb}
and showed the shortest time variations of its $\gamma$-ray emission
levels observed so far with doubling and decaying times in the energy
domain above $\sim$200 GeV as short
as 15 minutes \cite{gaidos}. The amplitude ratios between
the discovery flux level and flux maxima during short outburts
have been as large as 30, i.e., between 0.3 and 10 Crab,
where 1 Crab is the (time-independent) flux
of the standard candle of $\gamma$ astronomy, the Crab nebula.

BL Lac objects and flat spectrum radio-loud quasars (FSRQs)
are collectively called "blazars". The predominantly non-thermal
emission shows violent variability in most energy bands and superluminal motion
is observed in VLBI radio surveys.
The non-thermal emission is believed to originate in relativistic jets
oriented with small angles towards the line of sight.
The emission levels and time scales observed in the TeV domain, e.g., for
Mrk 421, clearly show that {\it relativistic beaming} has to be operational
in the sources, corroborating this model picture. The ultimate
energy source is believed to be gravitational, i.e., originating from
accretion of matter onto a supermassive ($\cal{O}$(10$^8$ - 10$^9$) \Msol)
black hole in the centre of the AGN. How the jets are formed
or how they are fueled is not yet understood. One viable model
is based, e.g., on the extraction of the rotational energy from the ergosphere
of the ($\sim$maximally) spinning black hole \cite{blandford}.
Irrespective of the actual formation and fueling of the jet
there are two models put forward to explain the observed radiation
from $\gamma$-ray emitting blazars. 

\begin{figure}[tbp]
\begin{center}
\psfig{figure=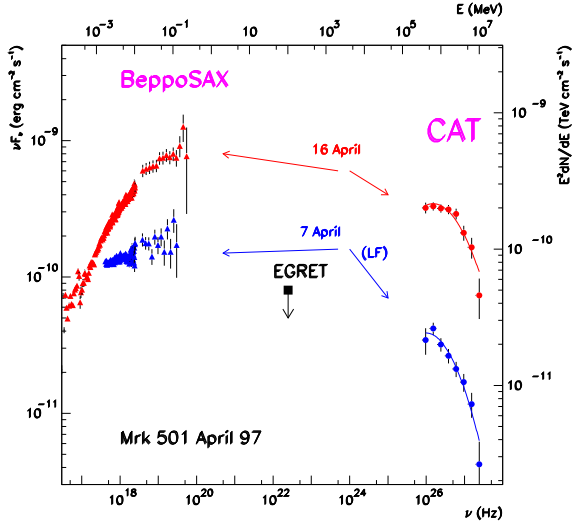,height=3.0in}
\end{center}
\caption{Mrk 501 X-ray and VHE spectra given as
$\nu F_\nu$. For April $7^\mathrm{th}$ and $16^\mathrm{th}$, BeppoS{\small AX}
data were taken from \cite{pian}.
The EGRET upper limit was taken from \cite{samuelson}
and corresponds to observations between
April $9^\mathrm{th}$ and $15^\mathrm{th}$. From \cite{djannati-atai}.}
\label{fig:gr6}
\end{figure}

The spectral energy
distributions of blazars in general seem to consist
of two parts. First, a low energy component rising from the radio up to a broad peak
between the infrared (IR) and X-ray wavelengths (see, e.g., figure~\ref{fig:gr6})
depending on the specific blazar type. This component is generally believed
to stem from incoherent synchrotron emission by relativistic electrons in the jets.
The origin and composition of the high energy component, however,
is still a matter of debate. Most popular is the model in which the
$\gamma$-rays are produced in optically thin
regions through inverse Compton scattering 
of low energy photons by the same electron population that
produces the synchrotron emission. Different models of this type differ,
e.g., in the origin of the low energy seed photons.
These photons can be synchrotron photons (Synchrotron Self Compton Model),
photons from the accretion disk, or other photons in the vicinity of the jet
(External Compton Models) (e.g., \cite{hoffman}). 
In a different type of model VHE $\gamma$-rays are
produced by proton-initiated cascades (PIC) \cite{mannheim}.
In this case normal plasma is injected into the jet and
shocks within the jet accelerate protons and electrons to very high energies
at which pions can then be photoproduced. All or parts of the VHE $\gamma$-ray emission in
this case is the end product of an electromagnetic cascade
developing in an optically thick acceleration region.
Note that this model was the only one {\it predicting} $\gamma$-ray
emission to energies up to and beyond 20 TeV. 
A common denominator of all models is that the $\gamma$-rays
originates close to the central black hole. Should the low energy fields
near to the accretion disk, however, be very intense the 
direct escape of the $\gamma$-rays
becomes difficult due to $\gamma \gamma$ $\rightarrow$ e$^+$e$^-$
pair production losses \cite{protheroe2}. VHE $\gamma$-ray data on time variability
and energy spectra taken concurrently with data in other energy bands
will help to constrain or rule out the models.

\begin{figure*}
\psfig{figure=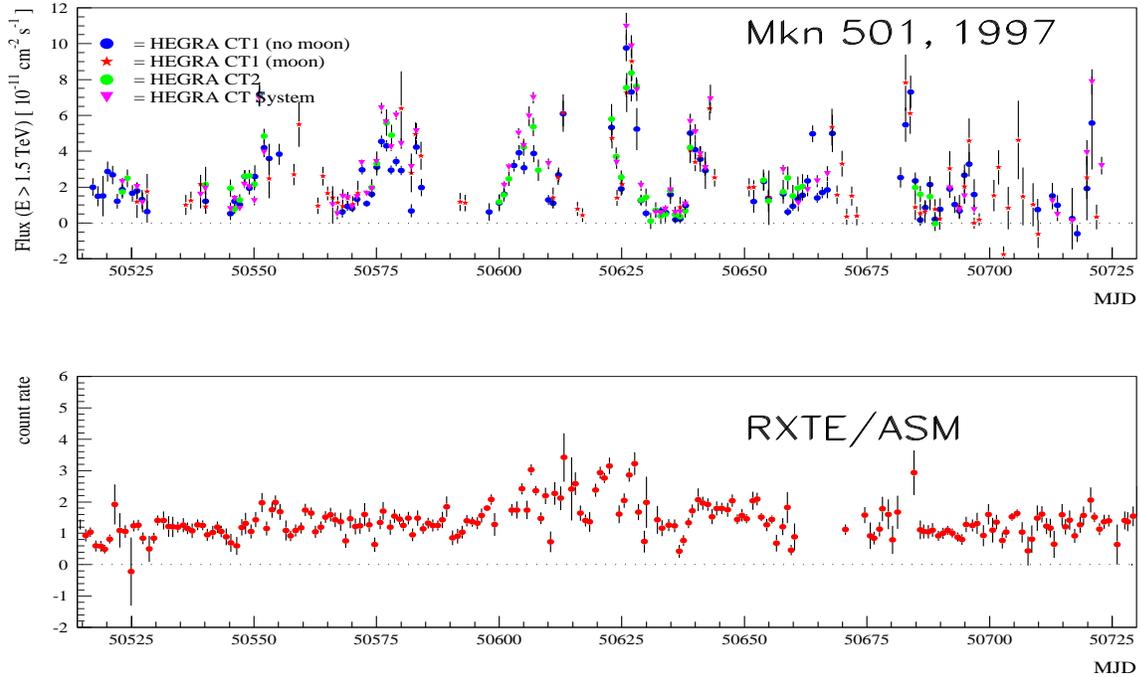,width=6in,height=3.8in}
\vspace{-10pt}
\caption{Light curves of Mrk 501 in 1997. Top: VHE $\gamma$-ray ($>$ 1.5 TeV)
emission measured by the HEGRA IACTs.
Modified Julian Day 50449 corresponds to January 1, 1997. 
Bottom: X-ray light curve measured by the RXTE ASM 
for 2 keV $\leq$ E $\leq$ 10 keV. From \cite{aharonian}.}
\label{fig:gr2}
\end{figure*}

The second extragalactic source, Mrk 501, was discovered in 1995 \cite{quinn}
as a weak source (0.08 Crab).
In the beginning of 1997 it showed a dramatic outburst at TeV energies which
lasted for the full 1997 observation period (March - September)
and was characterized by wildly varying emission levels.
The most complete light curve at energies larger than 1.5 TeV was
obtained by the 6 IACTs of the HEGRA collaboration who
operated at that time a stereoscopic system of 4 IACTs and 2 additional
stand-alone telescopes. In order to achieve this
good time coverage of Mrk 501 for the first time measurements during
moon time were performed 
with one of the stand-alone telescopes (CT1) \cite{kranich}. 
This observation mode considerably increased the
time coverage with only a moderate increase in energy threshold.
The light curve measured by all HEGRA IACTs 
is depicted in figure~\ref{fig:gr2} (top diagram).
Concurrently with HEGRA and many other IACTs (e.g., \cite{protheroe})
the Allsky Monitor (ASM) of the RXTE X-ray satellite took data on Mrk 501 in the
2 to 10 keV energy band. The corresponding light curve 
is also shown in figure~\ref{fig:gr2} (bottom diagram). 
HEGRA performed a correlation
analysis and found a moderate (0.611 $\pm$ 0.057) but clear (8.5 $\sigma$) 
correlation between the TeV and keV emission levels \cite{aharonian}
indicating a common emission region of the radiation.

The determination of the velocity of this emitting region within
the jet also illustrates the impact of such concurrent observations.
The bulk Lorentz factors, $\gamma_{bulk}$, of the 
relativistically moving plasma can be determined in two independent ways.
From radio observations with the Space-VLBI technique for Mrk 501
the inclination of the jet axis with the line of sight
was measured to be $\Theta$ $\sim$ 10$^{\circ}$ - 15$^{\circ}$.
At the same time the bulk velocity of the synchrotron emitting plasma, $\beta$,
was measured to be in the range from 0.990 to 0.999.
These numbers can be translated into a Lorentz factor, $\gamma_{bulk}$, between 7 and 22
or a Doppler factor $\delta$ = $\left[\gamma_{bulk} (1 - \beta\cos \Theta)\right]^{-1}$ 
between 1.3 - 5.6 \cite{giovannini}. The other method is based on the 
simultaneous observation of synchrotron spectra in the
optical/UV domain and TeV $\gamma$-ray spectra 
and makes use of the Synchrotron Self Compton model.
In this case the concurrent observations require $\delta$ $\gsim$ 5
\cite{buckley} consistent with the radio determinations.

HEGRA investigated the X-ray and TeV light curves
shown in figure~\ref{fig:gr2} for quasi-periodic oscillations (QPOs) \cite{kranich2}.
In the QPO analysis evidence for a 23 day period was found
in both light curves. In order to assign a statistical significance
a shot noise model assuming as null hypothesis independent, randomly
distributed flares in accordance with observations
was developed \cite{dejager}. 
The combined probability for
the 23 day period was thus determined to be $\cal{P}$ = 2.8 $e$-04 or 3.5$\sigma$.
Note that this analysis was only possible because the data taken during moon time
were available. Leaving out these data introduces large time gaps
and the observed QPO signal will be
washed out due to the introduction of aliasing effects.
The interpretation of the observed 23 day period is complicated by the
uncertainty of the exact location and environment of the emission region.
It could be due gravitomagnetic precession 
or g-mode oscillations of the accretion disk in analogy
to models put forward to explain 
QPOs observed during X-ray flares of the galactic analogon 
to AGN, the X-ray binaries \cite{cui}.
Or it could be due to the action of a close binary system of supermassive
black holes in the centre of AGN which is plausible based
on the heavy galaxy merger activity, e.g., observed in the Hubble
Deep Fields \cite{hdf}.
Other possibilities are radiating clumps on helical trajectories
or oblique shock fronts.

\begin{figure}[tbp]
\begin{center}
\psfig{figure=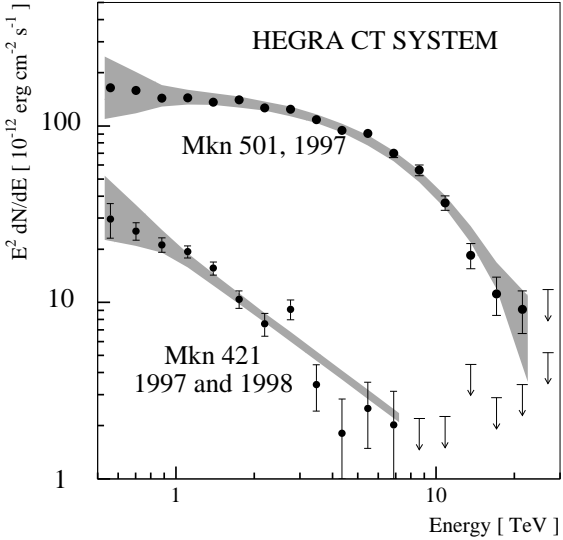,height=2.8in}
\end{center}
\caption{The energy spectra of Mrk 421 (1997 and 1998 combined) and Mrk 501 
(1997) as measured by the system of HEGRA IACTs. From \cite{aharonian2}.}
\label{fig:gr9}
\end{figure}

\begin{figure}[tbp]
\begin{center}
\psfig{figure=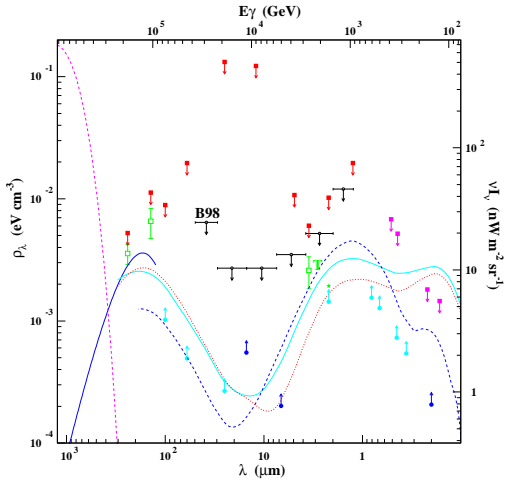,height=2.5in}
\end{center}
\caption{The density of the EBL between the far-infrared
and the ultraviolet. Filled squares with arrows
indicate 95\%CL upper limits by various experiments.
Open squares are detections (140 and 240 $\mu$m)
and a tentative detection (3.6 $\mu$m, denoted by T) by DIRBE.
The filled circles are lower limits from IRAS and ISO galaxy counts.
For references see \cite{vassiliev}.
The horizontal bars with arrows indicate the upper limits
from the Mrk 501 spectrum derived in \cite{biller}.
The dashed line above 300 $\mu$m indicates the level of the 2.7K CMB.
The dotted and dot-dashed lines are model predictions
from \cite{primack} and the dashed line a prediction
taken from \cite{malkan}. From \cite{vassiliev}.}
\label{fig:gr12}
\end{figure}

Besides the light curve the other important measurement
concerns the $\gamma$-ray energy spectra of AGN during high and low states.
From their observations of Mrk 501 in 1997 the CAT collaboration 
determined energy spectra for different $\gamma$-ray intensities \cite{chounet}.
Figure~\ref{fig:gr6} shows an example of the multi-wavelength spectra
of Mrk 501 as measured by the BeppoSAX X-ray satellite and the CAT telescope
on April 7$^{th}$ and April 16$^{th}$. 
The flux
in the $\gamma$-ray band differs by about a factor of 10. In addition
CAT found that the increase in flux was also accompanied by an increase in the spectral
hardness \cite{djannati-atai}. At somewhat higher energies this trend could
not be found in the HEGRA data \cite{aharonian2}.
The most striking aspect of the HEGRA energy spectrum of Mrk 501
in 1997, besides its independence from the flux level, is the
maximum energy observed from this source.
This energy spectrum together with the energy spectrum determined
for Mrk 421 from the combined 1997 and 1998 data
is shown in figure~\ref{fig:gr9}.
The 95\% CL lower limit on the maximum detected energy
from this extragalactic source is 16 TeV!
Due to the distance of Mrk 501 of about 500 million light years
the measurement of this maximum energy is a cosmologically
important signal. The reason for this is that
the TeV $\gamma$-rays have to traverse the 
diffuse extragalactic background light (EBL)
in order to arrive at the location of our Galaxy.
The EBL constitutes an important cosmological
signal as it integrates over the emission history
of the Universe on Hubble time and length scales.
Important parts of the EBL spectrum, amongst others
the infrared region, however, are
still only known very poorly from direct measurements
by satellite detectors.
TeV $\gamma$-rays can interact
with infrared photons 
to produce electron-positron pairs.
A large density of diffuse infrared photons 
therefore results in measurable absorption
features in extragalactic $\gamma$-ray spectra. As the energy
dependence of the EBL flux is roughly $\propto$ E$^{-1}$
{\it quasi-exponential} cutoffs of the TeV spectra are expected. 
By turning this argument around one sees that a beam of 
TeV $\gamma$-rays traveling cosmological
distances can be used to {\it probe} this background field. 
As the relevant Thomson cross section is
strongly peaked at the threshold this probing can
even be performed spectroscopically by observing
sources at different distances. Because of the energy
dependence of the EBL spectrum the cutoff
condition, i.e., optical depth reaching unity,
for sources at different cosmological distances occurs
at different $\gamma$-ray energies.
A 'cosmological $\gamma$-ray horizon with the Universe
opening up as one goes towards lower $\gamma$-ray energies
should result. 
A small fraction of the EBL spectrum is shown in figure~\ref{fig:gr12}.
What is striking is that in the shown
energy interval between the far infrared ($\sim$200 $\mu$m)
and the ultraviolet (UV) ($\sim$0.1 $\mu$m)
the figure is dominated by wildly scattered upper and lower limits.
The main reason for this is the enormous foreground radiation in this
spectral range by the zodiacal light in our solar system.
Besides the limits and measurements obtained with 'direct' methods
also the upper limits on the IR EBL level derived from the TeV $\gamma$-ray
spectrum of Mrk 501 as measured by the Whipple collaboration is shown
in figure~\ref{fig:gr12}.
A similar analysis was performed on the basis of the HEGRA data
resulting in an upper limit of 1.1 $\times$ 10$^{-3}$ eV/cm$^{3}$
at $\sim$25 $\mu$m {\it assuming} a specific shape of the
IR background \cite{funk}. As visible in figure~\ref{fig:gr12}
the application of the methods of TeV $\gamma$-ray astronomy
have already led to an improvement of the upper limits
in the mid-infrared range by more than an order of magnitude.

Especially in the mid-infrared range where the models predict a pronounced dip in the
diffuse extragalactic photon flux the dominating foreground will
also in the future probably not allow a certain direct determination of the 
flux level by infrared satellites. 
The level of the EBL in this range, however,
is an important signal of structure
formation in the Universe and constitutes a (complicated) convolution
of star formation rate, initial mass function and its evolution,
and the dust history of the Universe.
It is therefore of paramount interest
to determine the flux with the methods of TeV astronomy. As this means
measuring cutoff energies in the TeV domain, it makes it necessary to find
more close-by (on cosmological scales) TeV sources, i.e., raising the
instruments sensitivities as planned for the next generation
of $\gamma$-ray detectors.

The final result from TeV $\gamma$-ray astronomy reported on here
concerns the test of Lorentz invariance made possible by the combination
of high energy emission and cosmological distances.
Some of the ans\"atze for Quantum Gravity (QG), e.g., \cite{amelino}, 
yield non-Lorentz-invariant terms which lead to modified laws of propagation
and interaction of neutral
particles as a result of interactions with the quantum gravity medium.
Measurable time delays can be expected if the particles
have energies close to the QG scale (expected to be close
to the Planck scale, i.e., 10$^{19}$ GeV)
or if they have traversed cosmological distances.
Biller et al. \cite{biller2} analysed  
the shortest flare of Mrk 421 observed on May 15 1996 
by the Whipple collaboration \cite{gaidos}
for a time delay between $\gamma$-rays with energies less than 1 TeV and 
those with energies above 2 TeV. No difference was found within the 
measurable shortest time bin of 280 s and a lower limit of
4 $\times$ 10$^{16}$ GeV could be placed
on the relevant QG energy scale.
In general this type of investigation can be used to place
limits on any non-Lorentz invariant propagation term. By
searching for effects of this kind in very intense
transient phenomena occurring at cosmological distances like, 
e.g., GRBs, the upcoming MAGIC telescope will have sensitivity
at the Planck scale \cite{magnussen}. 
For comparison, the current best limits are about 1\% of the Planck scale.
 
\section{Neutrino Astronomy}
\label{sec:na}

The observation of low energy astrophysical neutrinos 
from the sun and SN1987A has led to important astrophysical
and particle physics results. The fundamentally new aspect of performing
neutrino astronomy in contrast to astronomy with electromagnetic
radiation lies in the penetration power of neutrinos, i.e.,
also optically thick regions can be studied in the neutrino 'light'. 
In addition neutrinos do not suffer attenuation losses
due to absorption on background radiations as do $\gamma$-rays or
ultrahigh energy hadrons.
We know from the observation of CRs that large numbers of hadrons are
accelerated to very high energies in astrophysical environments.
Whenever processes accelerate hadrons to VHE energies besides the
production of $\gamma$-rays we naturally expect neutrino fluxes
as least as large as the $\gamma$-ray fluxes and neutrino
and $\gamma$-ray astronomy are complementary approaches to
many astroparticle physics questions.
Also several astrophysical events, e.g., supernovae (SN)
or GRBs, are predicted to emit
their maximum power in the neutrino channel.
In summary, all environments where photoproduction of pions is likely
to occur are potential high energy neutrino sources.
These are the (unknown) sources of CRs,
the jets of AGN if protons are accelerated
within the jets, the Galactic disk because of CR interactions with 
the interstellar matter, or the centres of galaxy
clusters where UHECRs might be present.
In addition there are exotic sources like the 
pairwise annihilation of neutralinos 
and radiation
from topological defects. Note that the detection of
AGN as point sources for high energy neutrinos is
considered to be the {\it experimentum crucis} in distinguishing
between jet models where only electrons and positrons or where
'normal plasma', i.e., electrons and protons 
produce the observed VHE $\gamma$ radiation.
Although the neutrino-proton cross section
rises linearly up to energies of $\sim$ 10$^8$ GeV
before flattening due to $W$ propagator effects
the strongly falling fluxes expected from most conceivable sources
demands that high energy neutrino detectors have to be huge instruments.
For more information and the potential of this exciting new field see, e.g., 
\cite{gaisser2,halzen}.

The instruments that are taking data (BAIKAL, AMANDA-B13),
are under construction (AMANDA-B, ANTARES Phase II), or are in the 
test phase (NESTOR) are huge detectors
with effective collection areas between 2 $\times$10$^3$
and $\sim$10$^5$ m$^2$. Due to the expected low rates,
however, all these detectors are considered to be first or
second steps towards the construction of one or two
1 km$^3$ detectors (e.g., ANTARES, ICECUBE, NESTOR).
Basically the detectors are muon detectors
aimed at measuring $\nu_{\mu}$ and $\bar{\nu_{\mu}}$ induced 
{\it upward going} high energy muons. 
At high energies this gives the added advantage
of enlarging the fiducial volume of the detectors due to the
long range of high energy muons in water or ice.
Water or ice are the detector media of choice because the
detection of the muons utilizes the Cherenkov light emitted
by the water or ice when traversed by relativistic particles
with energies above the Cherenkov threshold.
The main parameters of these neutrino telescopes are
the angular and energy resolution. In general water detectors
will have a better angular resolution
but worse energy resolution compared to ice detectors.
In addition $\nu_{e}$ and $\bar{\nu_{e}}$ fluxes
can be easily measured at large energies based on the detection of
induced electromagnetic showers.
The major background to extraterrestrial neutrinos
is given by the {\it downward going} atmospheric muons and the 
{\it upward going} muon flux induced by atmospheric neutrinos
both of which are produced by the interaction of CRs in the earths atmosphere.
In addition to constituting two sources of background, however,
these two beams can be used as test beams for proving the
feasibility of these detectors by measuring the fluxes and angular
distributions of these beams at very high energies.

\begin{figure}[tbp]
\begin{center}
\psfig{figure=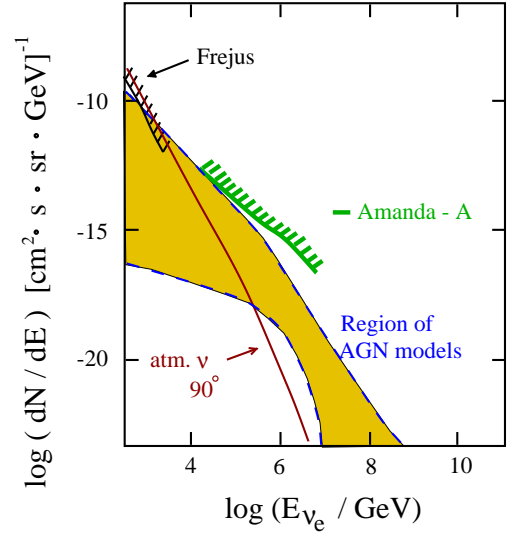,height=3.0in}
\end{center}
\caption{Limits on the diffuse $\nu_{e}$ + $\bar{\nu_{e}}$ flux 
as determined by the Frejus \cite{rhode} and AMANDA-A detectors.
From \cite{porrata}.}
\label{fig:na5}
\end{figure}

\begin{figure}[tbp]
\begin{center}
\psfig{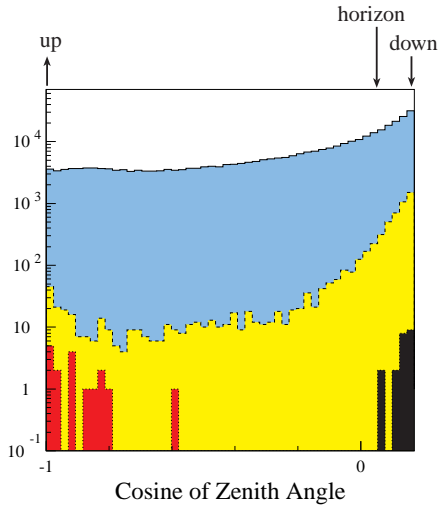}
\end{center}
\vspace{-10pt}
\caption{Reconstructed zenith angle distributions of 113 days of AMANDA-B10
data after quality cuts tightening from top to bottom.
The 17 events with a cos$\theta_{zenith}$ close to -1 are 
atmospheric neutrino candidates. From \cite{spiering}.}
\label{fig:na6}
\end{figure}

An ice detector currently under construction is the AMANDA-B detector
at the South Pole. It will have an effective collection area of 10,000 m$^2$,
an energy threshold of $\sim$50 GeV and an angular resolution
of about 2.5$^{\circ}$ per muon track \cite{halzen}.
While being constructed at different depths the AMANDA detector has 
been taking data. The AMANDA-A detector was installed
in a depth between 810 and 1000 m below the south pole's surface
and is characterized by a high concentration of residual
air bubbles leading to strong scattering of light.
While this dramatically worsened the angular resolution
it improved the calorimetric performance and 
the 1995 data cascade trigger data of AMANDA-A could be used
to put an upper limit on the diffuse flux of $\nu_{e}$ + $\bar{\nu_{e}}$
in the energy range from 10$^4$ to 10$^7$ GeV
\cite{porrata}. Due to the small size of AMANDA-A
this limit is, however,  not yet
restricting the model space of fluxes
predicted in AGN models (see figure~\ref{fig:na5}).
After AMANDA-A the installation towards the AMANDA-B detector
started with the deployment of 4 strings at a depth between 1.5 and 2 km.
For results from this detector see \cite{andres}.
The data taken 
during the first year of operation of the 10 string AMANDA-B10 detector installed in the depth
between 1500 and 1980 m
were analyzed for upward going muon candidates \cite{spiering}.
In a preliminary analysis of data taken during 113 days 17 events passed all selection cuts.
This corresponds to a reduction of the number of events
in the solid angle where the 'pencil like' AMANDA-B10
detector is sensitive by more than a factor of 10$^5$.
The result was found to be in accordance with the Monte Carlo expectation of
21.1 events.
The reconstructed zenith angle distributions for the full data set (4.9$\times$10$^8$ events),
and the data sets after different selection cuts are shown in figure~\ref{fig:na6}.

\begin{figure}[tbp]
\begin{center}
\psfig{figure=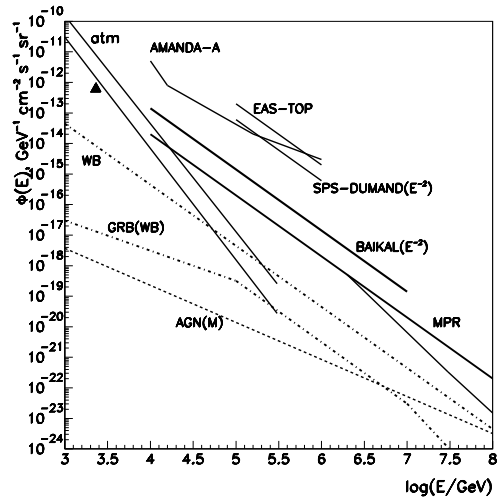,height=2.8in}
\end{center}
\caption{
Upper limits on the differential flux of high
energy neutrinos obtained by
different experiments and some upper bounds
on neutrino fluxes for different models.
For full references see \cite{balkanov}.
Dot-dash curves
labeled WB and GRB(WB) - upper bound and neutrino intensity
from GRB estimated by Waxman and Bahcall \cite{waxman2};
dashed curve labeled AGN(M) - neutrino intensity from AGN 
(Mannheim, model A \cite{mannheim2});
solid curves labeled MPR - upper bounds for $\nu_{\mu}+\bar{\nu_{\mu}}$
in \cite{mannheim2} for pion photo-production neutrino sources
with different optical depth $\tau$. 
The triangle denotes the limit obtained by Frejus
for an energy of 2.6 TeV :  $7 \cdot 10^{-13} \mbox{cm}^{-2}
\mbox{s}^{-1} \mbox{sr}^{-1} \mbox{GeV}^{-1}$ \cite{rhode}.
From \cite{balkanov}.}
\label{fig:na7}
\end{figure}

The BAIKAL detector, operational since April 1993, also took data
while being enlarged to its final NT-200 configuration in April 1998.
BAIKAL can be considered to be the water detector which proved
the feasibility of very high energy neutrino telescopes 
when it reconstructed the first two atmospheric neutrino candidates
from the initial NT-36 data taken between April 1994 and March 1995 \cite{spiering2}.
In its NT-96 configuration it took data from April 1996 until March 1997
and a  very tight limit could be placed on the diffuse $\nu_e$+$\nu_{\mu}$ flux 
for neutrino energies larger than 10 TeV \cite{balkanov}.
This limit together with some other experimental limits and 
model predictions is shown in figure~\ref{fig:na7}. Although the
BAIKAL NT-96 limit is still about a factor of 10 above the
highest model predictions it demonstrates the capability of
the underwater technique.
This technique is also going to be applied by
the ANTARES project planning the construction of
a 1 km$^3$ detector in the Mediterranean Sea (e.g.,
\cite{bertin}). Since 1997 ANTARES is conducting an intense
R \& D program aimed at the installation of a
0.1 km$^2$ detector in ANTARES Phase II by 2002 to 2003. 
In this approved phase 13 strings with $\sim$ 1000 PMTs
will be deployed in about 2500 m depth. Monte Carlo
simulations of the performance look very promising, e.g.,
an angular resolution of $\sim$0.2$^{\circ}$ is indicated.

\section{Outlook}
\label{sec:ol}
The major potential of Astroparticle Physics 
is given by the synergistic effects of combining
particle, nuclear and atomic physics with astrophysics and cosmology
and it has started to make an impact on important physics questions.
The improvement of detector technology coupled with
improved 'calibrations' of astrophysical beams, e.g.,
CR fluxes or $\gamma$ and $\nu$ fluxes from extragalactic objects,
will lead to an even larger sensitivity to fundamental
questions in the future.

\end{document}